\documentstyle[prb,aps,epsf]{revtex}
\begin{document}
\title{Density Waves in a Transverse Electric Field
}
\author{
Gilles Montambaux\\
{\it Laboratoire de Physique des solides, Associ\'e au CNRS, Universit{\'e}
Paris-Sud, 91405 Orsay, France}\\ }
\twocolumn[
 \maketitle
\widetext
\begin{abstract}
In a quasi-one-dimensional conductor with an open Fermi surface, a Charge or
a Spin
Density Wave phase can be destroyed by an electric field perpendicular to the
direction of high conductivity. This mechanism, due to the breakdown of
electron-hole symmetry, is very similar to the
orbital destruction of superconductivity by a magnetic field, due to
time-reversal symmetry.
\end{abstract}
                                                                                
]
\narrowtext
It is well-known that  superconductivity
becomes unstable in the presence of an external magnetic field
and that the transition temperature $T_c$ is reduced.
This is due to the breakdown of time-reversal symmetry which leads to
orbital frustration. More precisely, in the presence of a vector potential
${\bf A}$, a single electron wave function gets a phase shift
$ {e \over \hbar} \int {\bf A} d {\bf l} = {e \over \hbar} \int {\bf A}
{\bf v} d t$ and a  Cooper pair acquires
 a phase shift $ {2 e \over \hbar}\int {\bf A} d {\bf l}$. This
dephasing of the Cooper pair is at the origin of the reduction of $T_c$
\cite{gorkov,abrikosov}.

It is also known that this magnetostatic effect
has an electrostatic analog. An electron moving in a static potential $V$
acquires a phase $  {e \over \hbar} \int V  d t$. This is the electrostatic
Aharonov-Bohm effect\cite{AB}. Although the sign of this phase is not
changed with the velocity, it is changed with the sign of the carrier. That
is why  one expects
that an {\it electron-hole} pair will get a phase shift $ {2 e \over \hbar}
\int V d t$ exactly like the Cooper pair gets the phase $ {2 e \over \hbar}
\int {\bf A} d {\bf l}$ . In other words, the magnetic field breaks the
time reversal symmetry and the scalar potential breaks the electron-hole
symmetry. We show in this letter that, in an appropriate geometry,
this dephasing of the {\it electron-hole} pair may lead to the suppression
of the Charge Density Wave (CDW) or Spin Density
Wave (SDW) state
 by an  applied electric field. This field destroys the DW
 ordering in a very similar way that the magnetic field
destroys superconductivity (fig. 1).

The physical properties of CDW's and in particular
their elastic properties have been extensively studied. The main goal was
to describe the effect of an electric field {\it parallel} to the chains
and transport properties (like non-linear transport or conversion of
current at the boundaries)\cite{review,references}.
The  effect described here is specific of a quasi-1D system,
where the hopping integral $t_\perp$ between chains    plays                    
 an essential role. The Fermi surface is open                                   
and the electric field is applied along the direction {\it perpendicular} to
 the chains.                                                                    
 Although difficult to observe experimentally, this
new   effect  could
    be measurable in low $T_c$ Density Wave systems, for
example in the SDW state of Bechgaard salts.                                    
The effect described here is  purely thermodynamical.
We do not describe any effect related to nonlinear transport.

\begin{figure}[th]
\begin{center}
{\parbox[t]{4cm}{\epsfxsize 4cm
\epsffile{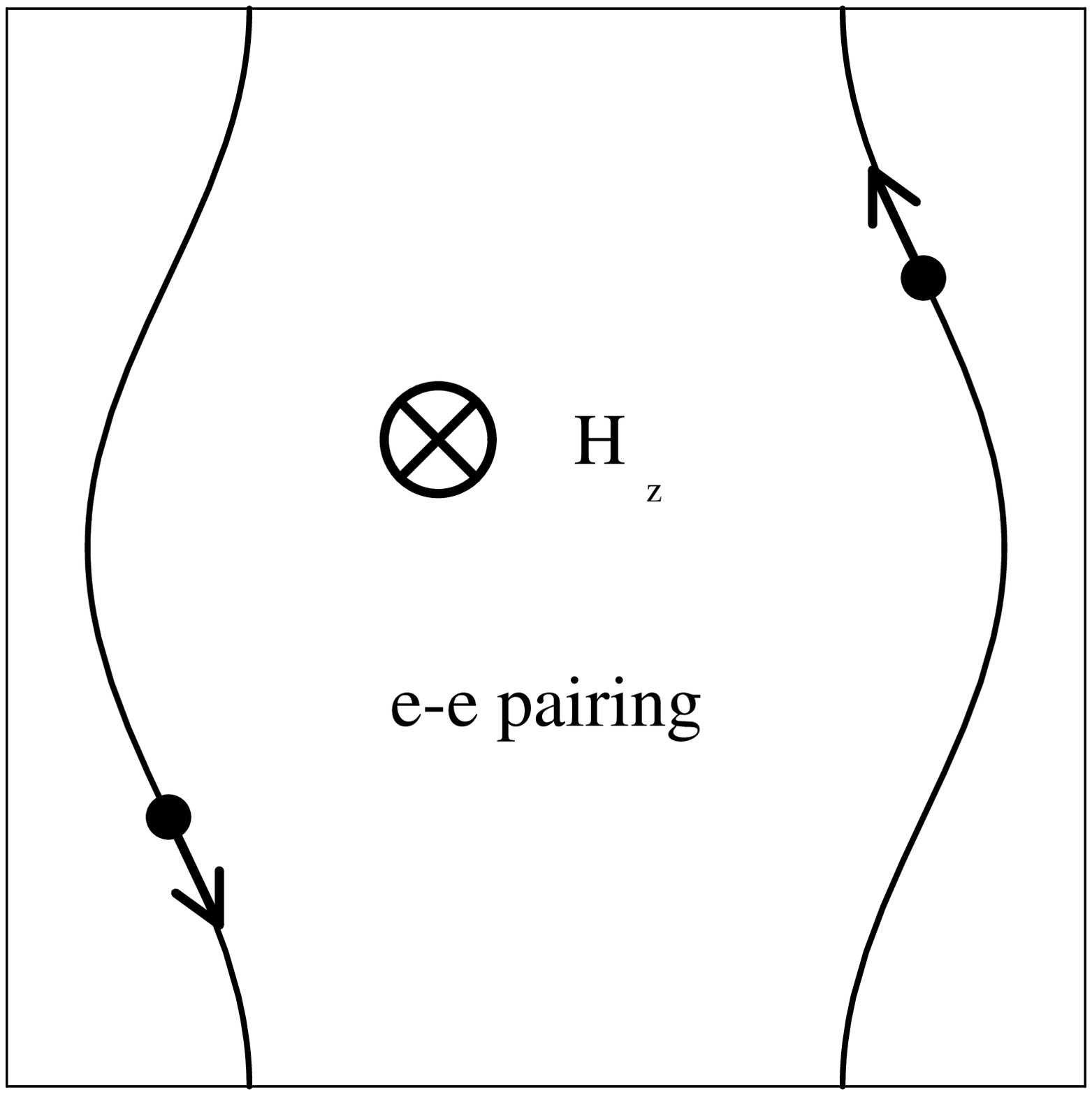}}
\hspace{-0cm} \parbox[t]{4cm}{\epsfxsize 4cm
\epsffile{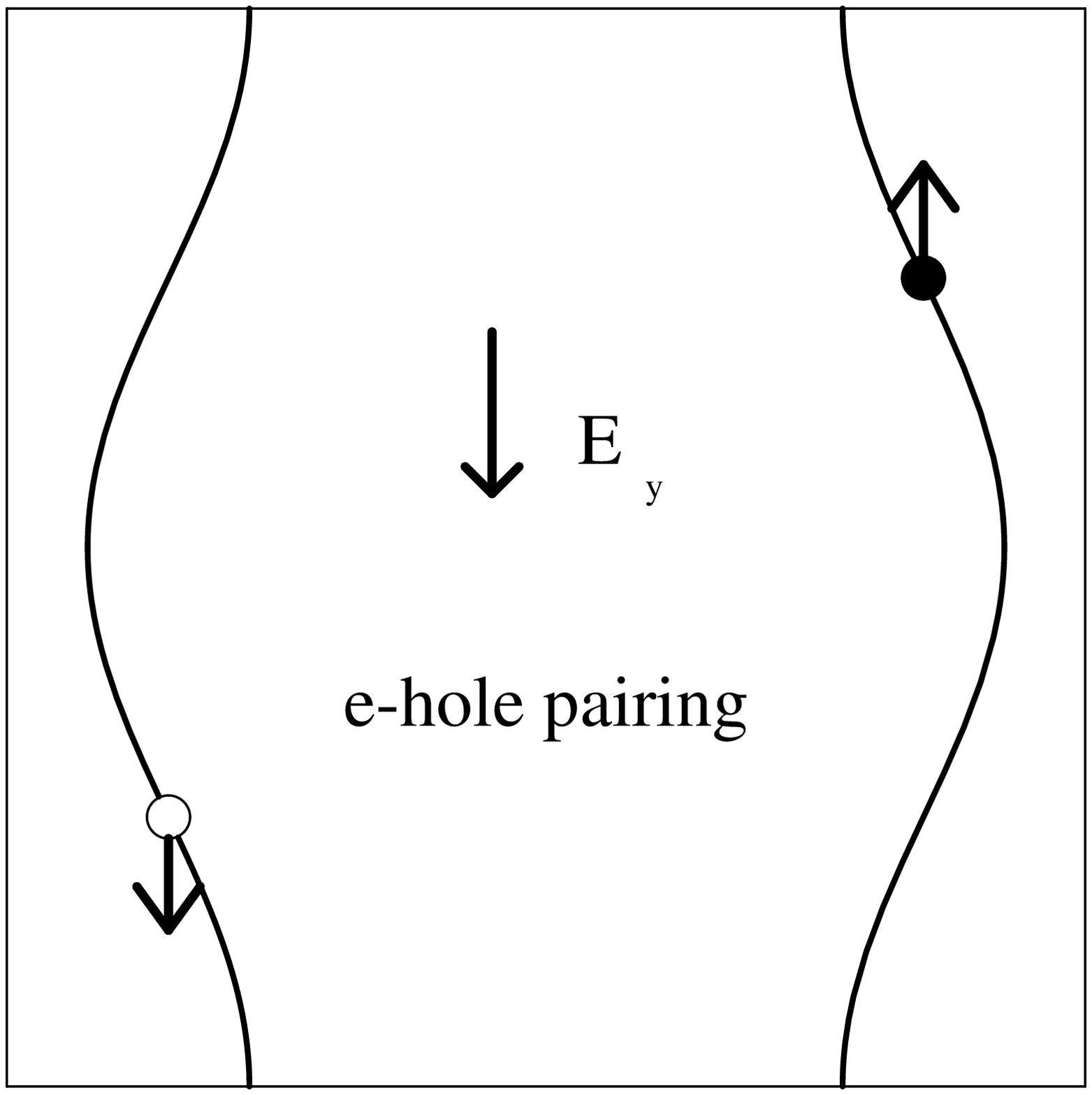}}
}
\end{center}
\caption{a)  In a magnetic field $H_z$, two electrons
move in opposite directions and accumulate opposite phases, leading to the
destruction of Cooper pairing. b)
 In a transverse electric field $E_y$, an
electron ($\bullet$) and a hole ($\circ$)
move in opposite directions and accumulate opposite phases, leading to
the  destruction of the electron-hole pairing. }
\label{fig} \end{figure}

Consider for
simplicity a two-dimensional
open Fermi surface, in a plane ($k_x,k_y$) . $k_x$ indicates the direction of
highest conductivity . An electric field $E_y$ is applied along the
$y$-direction. Using a tight-binding dispersion relation
linearized around the Fermi level in the x direction of high                    
conductivity, the Hamiltonian writes: ($\hbar = 1$ throughout the paper):
                                                                                
\begin{equation}                                                                
  {\cal H}({E_y})  =v_F(\vert k_x \vert -k_F)+t_\perp (k_y b
)- e E_y y
								  \label{HE}
\end{equation}                                                                  
                                                                                
\noindent
$v_F$ is the
Fermi velocity. The periodic function $t_\perp (p)$ describes                   
the warping of the Fermi surface. As an example, we will  take
$t_\perp (p) = -2t_b \cos (p)$ where $t_b$ is the transfer integral             
between chains. In such a case, the Fermi surface has perfect nesting at
wave
vector ${\bf Q}_0 = (2 k_F, \pi /b)$ so that, in zero electric field, a DW
is stable below
a critical  temperature $T_c$ given by a BCS form $T_c = T_c(E=0) = E_0 \exp
( -1 / g  N(\epsilon _F))$. $g$ is the coupling constant and
  $N(\epsilon _F)$ is the density of states at the Fermi level.

We now follow the lines of the usual semi-classical (eikonal) picture
used by Gor'kov\cite{gorkov} in the case of a magnetic field on a
superconductor.
 This approximation consists in replacing $y$ in eq.(\ref{HE}) by
its semi-classical zero field expression
\begin{equation}
y_{\alpha}(k_y,x) = \alpha {b t_{\perp}^{\prime}(k_y b) \over v_F} x
								 \label{SC}
\end{equation}

\noindent
$\alpha = \pm 1$ describes the two sides of the Fermi surface.
As a result, the wave function writes:

\begin{equation}
\phi_{{\bf k},\alpha}({\bf r}) = {1 \over \sqrt{S}}\exp[i  {\bf k r}
+ i \varphi (x,0) ]
\label{phi}
\end{equation}
\noindent
$S$ is the area.
 As explained in the introduction, the additional phase
factor:
\begin{equation}
\varphi (x,x') = i \alpha {e E_y \over v_F}
\int_{x'}^x  y_{\alpha}(k_y,x) dx
\label{phi1}
\end{equation}
\noindent
has indeed the form $ e \int V dt$ ($\hbar =1$) since $V = e E_y y$ and $d
t = d x/ v_F$. The Green function for an electron on the
right (left) side of the Fermi surface, written in a mixed
representation\cite{GL} is:
\begin{equation}
  G_{\alpha}(x,x^{\prime},k_y,\omega _n,E_y) =
  G_{\alpha}^0(x,x^{\prime},k_y,\omega _n)
e^{i \varphi(x,x')}
							   \label{greena}
\end{equation}

\noindent
$\omega_n$
are the Fermion Matsubara frequencies. $G_{\alpha}^0$ is the $E_y=0$ Green
function.

We now write the linearized
integral equation which describes the stability                                 
of the DW phase, at the wave vector ${\bf Q} = (Q_{x}, Q_y
)$ ,

\begin{equation}                                                                
 {\Delta({\bf Q},x) \over g} =
 \int                                                                           
 \nolimits dx' K({\bf Q},x,x') \Delta({\bf Q},x')
							  \label{gap}
\end{equation}
                                                                                
\noindent                                                                       
where, in the mixed representation, the kernel K of this
one-dimensional equation is given by:
\begin{eqnarray}
 K({\bf Q},x,x') =
   T \sum_{\omega_n >0}                                                         
 \int                                                                           
 \nolimits  {dk_y b \over {2 \pi}}
  G_+^0(x,x',k_y ,\omega _n) \times  \nonumber  \\
  G_-^0(x',x,k_y-Q_y,\omega _n)
  \exp[i Q_\parallel (x-x')]\exp [ i \phi]
							   \label{kernel}
\end{eqnarray}
                                                                                
\noindent
where
\begin{equation}
  \phi = { e E_y \over v_F}\int_{x^{\prime}}^x [y_+(k_y, x) +y_-(k_y
-Q_y, x)]dx
							   \label{phi2}
\end{equation}

\noindent                                                                       
This kernel has the same structure as the superconducting kernel in
a magnetic
field. The phase factor (\ref{phi2}) which has the form  $2 e \int V
dt$ plays a role
similar to the phase factor induced by the vector potential in the case of a
magnetic field $\exp[2 i e \int
{\bf A} d{\bf l}]$  ($\hbar = 1$).

Inserting eq. \ref{SC} into the phase factor (\ref{phi2}), performing
the Matsubara sum and the integral in eq. \ref{kernel} one finds that,
close to the nesting vector ${\bf Q}_0$, the gap equation writes:
\begin{eqnarray}
 {\Delta({\bf Q},x) \over g} =
 \int_{u>d}
  {du \over x_T\sinh({ u \over x_T})}   \times \nonumber \\
 J_0[{2t_b \over v_F} u (q_y + {2 e E_y b x \over v_F})]
 \Delta({\bf Q},x+u)
			\label{gap2}
\end{eqnarray}

\noindent
where $q_y = \pi /b - Q_y $.
$x_T= v_F/(2 \pi T)$ is the thermal length . $d$ is a lower cut-off such
that $1/g = \ln(\pi d T_c/v_F)$.
 In order to discuss the DW stability
criterion,
we first derive the linearized Ginzburg-Landau equation from the gap
equation  (\ref{gap},\ref{kernel})
 \cite{gorkov}. Expanding $\Delta(x+u)$
and the Bessel function $J_0$
for small separation $u$,
 it is then straightforward to get the equation
for the spatial variation of the order parameter                                
$\Delta({\bf r}) = \Delta \exp(- i {\bf Q}_0 {\bf r}) f({\bf r})$, where
$f= f({\bf r})$ obeys:

\begin{equation}                                                                
 [{1 \over 4} {\partial ^2 \over \partial x^2}
 + {t_b^2 b^2 \over  2 v_F^2}
 ({\partial \over \partial y} + {2 i e E_y x \over v_F})^2
 +{4 \pi ^2 T_c^2 \over 7 \zeta (3) v_F^2} (1-{T \over T_c})]f
=0                                                                              
							     \label{gl}
\end{equation}                                                                  
or,
by gauge transformation:

\begin{equation}
 [-{1 \over 4} ({\partial  \over i \partial x} +{2 e V
\over v_F})^2  + {t_b^2 b^2 \over  2 v_F^2}
 {\partial^2 \over i \partial y^2}
 +{4 \pi ^2 T_c^2 \over 7 \zeta (3) v_F^2} (1-{T \over T_c})] f
 =0
							     \label{gl2}
\end{equation}

This linearized Ginzburg-Landau equation is identical to the one which
describes the superconducting order parameter in a magnetic field.
The analogy with the magnetic field effect on superconductivity is  clearly
seen from the structure
of the Hamiltonian (\ref{HE}) which is very similar to the Hamiltonian in a
magnetic field ${\bf H} = (0,0,H_z)$ perpendicular to the plane.
 Using the Landau gauge
${\bf A} = (0,-H_zy,0)$, one gets

\begin{equation}
  {\cal H}(H_z)  =v_F(\vert (k_x + eH_z y)\vert  -k_F)+t_\perp cos(k_y
b)                                                                 \label{HH}
\end{equation}
The similar form of the two Hamiltonians clearly reflects the physical
description presented above. The absolute value
in eq.(\ref{HH}) simply reflects that, on the left side  of the Fermi
surface, the electron travels in the same direction in the presence of
$E_y$ or $H_z$, while, on the right side, it travels in opposite
directions (fig. 1).  This analogy between the two Hamiltonians is
 of course due to the quasi-1D structure of the Fermi surface
and originates from the linearization
of the dispersion relation along x.
Using the different gauge
${\bf A} = (0,H x ,0)$, it may be seen that the superconducting kernel
contains the phase factor

\begin{equation}
\exp[i \int_{x^{\prime}}^x t_\perp(p -eHbx) - t_\perp(p-q_y+eHbx) dx]
							      \label{supra}
\end{equation}

\noindent
Small $x$ expansion of the periodic factors $t_\perp$ directly leads to
an expression similar to eq.(\ref{phi2}).

From standard manipulation of eq.(\ref{gl2}) \cite{tiley}, we find that the
critical
temperature decreases linearly as $T_c(E) = T_c (1-E/E_c)$.
A simple extrapolation of
this linear behavior down to zero temperature\cite{valid} defines a critical
field $E_c$ at which the DW completely disappears. $E_c$ is given by
                                                                                
\begin{equation}                                                                
 2 e b E_c =                                                                    
  {8 \sqrt2 \pi ^2 T_c^2(0) \over 7 \zeta (3) t_b}
								 \label{Ec}
\end{equation}                                                                  
\noindent
This relation has a simple meaning. Noting that $T_c \propto \Delta$, the
critical field is such that $2 e E_y \xi_y \simeq \Delta$: the potential
energy of the electron-hole pair
 is of the order of the binding energy of the pair $\Delta$ (
$\xi_y \simeq {v_y \over \Delta} \simeq {t_b b \over
\Delta}$ is the transverse coherence length).
 Below $T_c$, the CDW order parameter should exhibit a non-uniform structure
similar to the mixed state of superconductors.

 Expression (\ref{Ec})
for the critical field shows that the destruction of the DW is a                
2D effect. This effect disappears when the coupling $t_b$
between chains vanishes.  The critical field is smaller when $t_b$ is large.
On the other hand, it should not be too large so
that the Fermi surface remains open and that a DW can still exist.

 The above derivation has been presented in the case of a perfect
nesting. An imperfectly nested DW can also been destroyed by an electric
field. We have checked that the GL equation has the same structure
in the case of imperfect nesting.

We see from eq. (\ref{Ec}) that the effect of the electric field is easier
to detect in DW systems with low critical temperature so that a small field
can be applied.
 Using the parameters of Bechgaard salts \cite{js}, one
can estimate
$E_c$ of the order of  $E_c = 25 T_c^2 V/cm$.
In order to have an electric field as small as possible, one should  use
a salt with a very small $T_c$.
 A favorable situation to observe this new effect is to apply
pressure                                                                        
on the SDW of a Bechgaard salt like                                             
$(TMTSF)_2 PF_6$, in order to reduce its critical temperature ($10K$ at
ambient pressure\cite{js}). The linear decrease of $T_c(E)$ could then be
observed in a typical electric field of order $10 V/cm$.

An experimental difficulty to observe this effect may be due to the
screening of the electric field in the metallic phase and along the
transition line so that rather than measuring directly the decrease of
$T_c(E)$ (by measures of the resistance R(T) at different fields for
example), it may be more appropriate to work at very low temperature well
below $T_c$ so that the DW order parameter is maximum and the
screening is minimum, and to apply the electric field.
Another possibility may be to apply a voltage to the sample with contacts so
that a small current will flow long the $y$ direction.

During the completion of this work, I have been aware of a preprint by M.
Hayashi and H. Yoshioka, who have treated a similar problem\cite{Hayashi96}.
\medskip

Acknowledgments: I benefited from useful discussions with
 H. Bouchiat, N. Dupuis, M. Gabay, P. Lederer, B. Reulet and  H. Schulz.

\newpage


\begin{thebibliography}{10}
\bibitem{valid} The Ginzburg-Landau expansion is valid  for $ebE \ll
T^2/t_b$
\end{thebibliography}
  \end{document}